\documentclass{aa}  
\usepackage{graphicx}
\usepackage{txfonts}
\usepackage[colorlinks=true,
			linkcolor=blue,
			citecolor=blue, 
			filecolor=blue, 
			urlcolor=blue]{hyperref}
\usepackage[switch]{lineno}
\usepackage{txfonts}
\usepackage{natbib}
\usepackage{graphicx, xspace}
\usepackage{multirow}
\usepackage{ amssymb }
\usepackage{subfigure}
\usepackage{epstopdf}
\usepackage{lineno}

\usepackage{natbib}
\bibpunct{(}{)}{;}{a}{}{,}

\begin{document} 

\title{Radio emission in Mercury magnetosphere}

%\subtitle{}
\titlerunning{Radio emission in Mercury magnetosphere}
\authorrunning{Varela et al.}

   \author{J. Varela\inst{1}
          V. Reville\inst{1}
          A. S. Brun\inst{1}          
          F. Pantellini\inst{2}
          and P. Zarka\inst{3}   
          }

   \institute{AIM, CEA/CNRS/University of Paris 7, CEA-Saclay, 91191 Gif-sur-Yvette, France \\
              \email{\href{mailto:deviriam@gmail.com (telf: 0033782822476)}{deviriam@gmail.com}}
         \and
             LESIA, Observatoire de Paris, CNRS, UPMC, Universite Paris-Diderot, Place J. Janssen, 92195 Meudon, France
          \and
              LESIA \& USN, Observatoire de Paris, CNRS, PSL/SU/UPMC/UPD/SPC, Place J. Janssen, 92195 Meudon, France \\
             }

\date{version of \today}

% \abstract{}{}{}{}{} 
% 5 {} token are mandatory
 
  \abstract
  {
     \textit{Context:} Active stars possess magnetized wind that has a direct impact on planets that can lead to radio emission. Mercury is a good test case to study the effect of the solar wind and interplanetary magnetic field on radio emission driven in the planet
magnetosphere. Such studies could be used as proxies to characterize the magnetic field topology and intensity of exoplanets. \\
     \textit{Aims:} The aim of this study is to quantify the radio emission in the Hermean magnetosphere. \\
     \textit{Methods:} We use the MHD code PLUTO in spherical coordinates with an axisymmetric multipolar expansion for the Hermean magnetic field, to analyze the effect of the interplanetary magnetic field (IMF) orientation and intensity, as well as the hydrodynamic parameters of the solar wind (velocity, density and temperature), on the net power dissipated on the Hermean day and night side. We apply the formalism derived by Zarka [2001, 2007] to infer the radio emission level from the net dissipated power. We perform a set of simulations with different hydrodynamic parameters of the solar wind, IMF orientations and intensities, that allow us to calculate the dissipated power distribution and infer the existence of radio emission hot spots on the planet day side, and to calculate the integrated radio emission of the Hermean magnetosphere. \\
      \textit{Results:} The obtained radio emission distribution of dissipated power is determined by the IMF orientation (associated with the reconnection regions in the magnetosphere), although the radio emission strength is dependent on the IMF intensity and solar wind hydro parameters. The calculated total radio emission level is in agreement with the one estimated in [Zarka, 2001], between $5 \times 10^{5}$ and $2 \times 10^{6}$ W.
    }


\maketitle
\tableofcontents

%\linenumbers
%\modulolinenumbers[5]

\section{Introduction}

An obstacle facing a magnetized flow leads to the partial dissipation of the flow energy. Part of the energy is dissipated as radiation in different ranges of the electromagnetic spectrum, depending on the incoming flow properties and the intrinsic magnetic field of the obstacle. This scenario describes the interaction of the stellar wind with the magnetosphere and atmosphere of planets or other stars. 

The power dissipated in the interaction of a magnetized flow with an obstacle can be sized as the intercepted flux of the magnetic energy ($[P_{d}] \approx B^2 v \pi R^{2}_{obs} / 2 \mu_{0}$), with $B$ the magnetic field intensity perpendicular to the flow velocity in the frame of the obstacle, $\mu_{o}$ the magnetic permeability of the vacuum, $v$ the flow velocity and $R_{obs}$ the radius of the obstacle. See \citet{Saur,Strugarek} for a description of the size and shape of the intersecting region and location of the maximal Poynting flux generation.

The planets of the solar system with intrinsic magnetic fields are emitters of cyclotron MASER emission at radio wavelengths, generated by energetic electrons (keV) traveling along the magnetic field lines, particularly in the auroral regions \citep{Wu}. The source is the reconnection region between the interplanetary magnetic field (IMF) and the intrinsic magnetic field of the planet, although in gaseous planets as Jupiter there are other internal sources like the plasma released from Io's torus or the fast rotation of the planet. The magnetic energy is transferred as kinetic and internal energy to the electrons (consequence of the local balance between Poynting flux, enthalpy and kinetic fluxes). Most of the power transferred is emitted as aurora emission in the visible electromagnetic range, but a fraction is invested in cyclotron radio emission \citep{Zarka5}. 

Radiometric Bode's law links up incident magnetized flow power and obstacle magnetic field intensity with radio emission as $[P_{rad}] = \beta [P_{d}]^{n}$, with $[P_{rad}]$ the radio emission and $\beta$ the efficiency of dissipated power to radio emission conversion with $n \approx 1$ \citep{Zarka3,Zarka4}. Recent studies pointed out $\beta$ values between $3 \cdot 10^{-3}$ to $10^{-2}$ \citep{Zarka}.

The power dissipated in the interaction between solar wind and magnetosphere field is strongly variable. Indeed several factors influence the nature of the interaction. The topology of the planet magnetic field is affected by the IMF orientation and intensity, as well as hydrodynamic parameter of the solar wind (SW) like density, velocity or temperature, leading to different distributions of radio emission hot spots and total emissivity. This is the case of Mercury, where the proportion of IMF and Hermean (Mercury) magnetic field intensity, defined as $\alpha = B_{sw}/B_{M}$ \citep{Baker,Baker2}, oscillates from 0.3 during a coronal mass ejection, $B_{sw} \approx 65 $ nT, to 0.04 for a period of low magnetic activity of the Sun, $B_{sw} \approx 8 $ nT \citep{Anderson,Johnson}. IMF orientations can depart further from Parker Spiral \citep{Parker}, leading to different configurations of the Hermean magnetosphere \citep{Fujimoto} due to the reconnection with the planet magnetic field \citep{Slavin}. At the same time, SW hydrodynamic parameters predicted by ENLIL (time-dependent 3D MHD model of the heliosphere) + GONG WSA (prediction of background solar wind speed and IMF polarity) cone models show too a large range of possible values: between $12 - 180$ cm$^{-3}$ for density, $200 - 500$ km/s for velocity and $2 \cdot 10^{4} - 18 \cdot 10^{4}$ K for temperature \citep{Odstrcil,Pizzo}. Consequently, a parameter study is required to analyze the radio emission from Mercury.

First measurements of the Hermean magnetic field by Mariner 10 identified a dipolar moment of $195nT*R^{3}_{M}$ ($R_{M}$ Mercury radius) \citep{Slavin2}, further refined by MESSENGER observations leading to a more precise description as a multipolar expansion \citep{Anderson2}. Electron cyclotron frequency in the Hermean magnetosphere is smaller than the plasma frequency of the SW, so the radio power in Mercury is expected to be trapped into the magnetosphere \citep{Zarka2}. 

The aim of this study is to calculate the radio emission driven in the interaction of the solar wind with the Hermean magnetosphere, analyzing the kinetic and magnetic energy flux distributions as well as the net power dissipated on the planet day and night side. The analysis is performed for different orientations of the IMF and SW hydro parameters. The radio emission from planetary magnetospheres is a source of information to foreseen the topology and intensity of exoplanets magnetospheres.

The interaction of the SW with the Hermean magnetosphere is studied using different numerical frameworks as single \citep{2008Icar..195....1K,2015JGRA..120.4763J}, multifluid \citep{2008JGRA..113.9223K} and hydrid codes \citep{2010Icar..209...46W,Muller2011946,Muller2012666,2012JGRA..11710228R}. The simulations show that the Hermean magnetic field is enhanced or weakened in distinct locations of the magnetosphere according to the IMF orientation, modifying its topology \citep{Slavin,2000Icar..143..397K,2009Sci...324..606S}. To perform this study we use the MHD version of the single fluid code PLUTO in spherical 3D coordinates \citep{Mignone}. The Northward displacement of the Hermean magnetic field is represented by a axisymmetric multipolar expansion \citep{Anderson3}. Present study is based in previous theoretical studies devoted to simulate global structures of the Hermean magnetosphere using MHD numerical models \citep{Varela,Varela2,Varela3}.

This paper is structured as follows. Section II, description of the simulation model, boundary and initial conditions. Section III, analysis of the radio emission generation for configurations with different IMF orientations and intensities. Section IV, study of the radio emission generation for configurations with different SW hydro parameters. Section V, conclusions and discussion.

\section{Numerical model}

We use the ideal MHD version of the open source code PLUTO in spherical coordinates to simulate a single fluid polytropic plasma in the non resistive and inviscid limit \citep{Mignone}.

The conservative form of the equations are integrated using a Harten, Lax, Van Leer approximate Riemann solver (hll) associated with a diffusive limiter (minmod). The divergence of the magnetic field is ensured by a mixed hyperbolic/parabolic divergence cleaning technique \citep{Dedner}.

The grid is made of $196$ radial points, $48$ in the polar angle $\theta$ and $96$ in the azimuthal angle $\phi$ (the grid poles correspond to the magnetic poles). The numerical magnetic Reynolds number of the simulations due to the grid resolution is $R_{m}= V L/\eta \approx 1350$, with $V = 10^{5}$ m/s and $L = 2.44 \cdot 10^{6}$ m the characteristic velocity and length of the model, and $\eta \approx 1.81 \cdot 10^{8}$ m$^{2}$/s the numerical magnetic diffusivity of the code. The numerical magnetic diffusivity was evaluated in dedicated numerical experiments using a model with the same grid resolution but a simpler setup.

In order to have a realistic representation of the Hermean magnetic field we make use of the current best knowledge of its topology and amplitude based on MESSENGER data. \citet{Anderson4} study provides a multipolar expansion of the field assuming an axisymmetric model, thus we implement in our setup an axisymmetric multipolar field up to $l = 4$ and $m = 0$. The magnetic potential $\Psi$ is expanded in dipolar, quadrupolar,
octupolar and hexadecapolar terms:

\begin{equation} \label{eq:1}
\Psi (r,\theta) = R_{M}\sum^{4}_{l=1} (\frac{R_{M}}{r})^{l+1} g_{l0} P_{l}(cos\theta)  
\end{equation}

The current free magnetic field is $B_{M} = -\nabla \Psi $. $r$ is the distance to the planet center, $\theta$ the polar angle and $P_{l}(x)$ the Legendre polynomials. The numerical coefficients $g_{l0}$ taken from Anderson et al. 2012 are summarized in Table 1.

\begin{table}[h]
\centering
\begin{tabular}{c | c c c c}
coeff & $g_{01}$(nT) & $g_{02}/g_{01}$ & $g_{03}/g_{01}$ & $g_{04}/g_{01}$  \\ \hline
 & $-182$ & $0.4096$ & $0.1265$ & $0.0301$ \\
\end{tabular}
\caption{Multipolar coefficients $g_{l0}$ for Mercury's internal field.}
\end{table}

The simulation frame is such that the z-axis is given by the planetary magnetic axis pointing to the magnetic North pole and the Sun is located in the XZ plane with $x_{sun} > 0$. The y-axis completes the right-handed system. 

The simulation domain is confined within two spherical shells centered in the planet, representing the inner and outer boundaries of the system. The shells are at $0.6 R_{M}$ and $12 R_{M}$. Between the inner shell and the planet surface (at radius unity in the domain) there is a "soft coupling region" where special conditions apply. Adding the soft coupling region improves the description of the plasmas flows towards the planet surface, isolating the simulation domain from spurious numerical effects of the inner boundary conditions \citep{Varela2,Varela3}. The outer boundary is divided in two regions, the upstream part where the solar wind parameters are fixed and the downstream part where we consider the null derivative condition $\frac{\partial}{\partial r} = 0$ for all fields ($\vec{\nabla} \cdot \vec{B} =0$ condition is conserved because at $12 R_{M}$ the IMF is dominant and constant). At the inner boundary the value of the intrinsic magnetic field of Mercury is specified. In the soft coupling region the velocity is smoothly reduced to zero when approaching the inner boundary. The magnetic field and the velocity are parallel, and the density is adjusted to keep the Alfven velocity constant $\mathrm{v}_{A} = B / \sqrt{\mu_{0}\rho} = 25$ km/s with $\rho = nm_{p}$ the mass density, $n$ the particle number, $m_{p}$ the proton mass and $\mu_{0}$ the vacuum magnetic permeability. In the initial conditions we define a paraboloid in the night side with the vertex at the center of the planet, defined as $r < 1.5 - 4sin(\theta)cos(\phi) / (sin^{2}(\theta)sin^{2}(\phi)+cos^{2}(\theta))$, where the velocity is null and the density is two order of magnitude smaller than in the solar wind. The IMF is also cut off at $2 R_{M}$.

We assume a fully ionized proton electron plasma, the sound speed is defined as $\mathrm{v}_{s} = \sqrt {\gamma p/\rho} $ (with $p$ the total electron + proton pressure), the sonic Mach number as $M_{s} = \mathrm{v}/\mathrm{v}_{s}$ with $\mathrm{v}$ the velocity.

The radio emission study for different IMF intensities is limited to the range of $10$ to $30$ nT. Lower or higher IMF intensities correspond to a solar magnetic activity below or over the average, not considered in the present communication. In the radio emission study with different hydro parameter of the SW and fixed IMF orientation and intensity, we adopt the maximum and minimum values expected by ENLIL + GONG WSA cone models for the SW density, temperature and velocity. The radio emission for configurations mimicking CME conditions are out of the scope of this study.

\section{Effect of the interplanetary magnetic field}
\label{IMF effect}

In this section we estimate the radio emission in Mercury for different orientations and intensities of the IMF. We calculate the power dissipated by the interaction of the SW with the Hermean magnetosphere on the planet day side and the reconnection region of the magnetotail on the planet night side. The effect of the reconnections on the day side is fully accounted for the study.The energy flux is calculated as a combination of kinetic $P_{k}$ (associated with the solar wind dynamic pressure) and magnetic terms $P_{B}$ (due to the reconnection between the IMF and the Hermean magnetic field):

\begin{equation} \label{eq:6}
P_{k} = \frac{1}{2}\rho \vec{\mathrm{v}} |\vec{\mathrm{v}}^{2}|
\end{equation}
\begin{equation} \label{eq:7}
P_{B} = \frac{\vec{E}\wedge\vec{B}}{\mu_{0}} = \frac{(\eta\vec{J} - \vec{\mathrm{v}}\wedge\vec{B})\wedge\vec{B}}{\mu_{0}}
\end{equation}
where $E$ is the electric field and $J$ the current density. The effect of the numerical magnetic diffusivity $\eta$ is negligible in the calculation of the power dissipation on the planet day side, but it is considered in the reconnection region.

The net power dissipated is calculated as the volume integral of $P_{k}$ and $P_{B}$:

\begin{equation} \label{eq:8}
[P_{k}] = \int_{V}  \vec{\nabla} \cdot \left(\frac{\rho \vec{\mathrm{v}} |\vec{\mathrm{v}}^{2}|}{2} \right) dV 
\end{equation}
\begin{equation} \label{eq:9}
[P_{B}] = \int_{V} \vec{\nabla} \cdot \frac{\vec{E}\wedge\vec{B}}{\mu_{0}} dV
\end{equation}

On the day side, the volume integrated extends from the bow shock to the inner magnetosphere (magnetosheath and magnetopause included). On the night side the integrated volume is localized around the X point of the magnetotail, covering the region where the intensity of the planet magnetic field is at least $10 \%$ smaller than for a configuration without IMF.

\subsection{Orientation of the interplanetary magnetic field}

We analyze first the effect of the IMF orientation, fixing the hydrodynamic parameters of the SW and the module of the IMF to $30$ nT. The hydrodynamic parameters of the solar wind in the simulations are summarized in Table 2. Figure 1 shows a 3D view of the system for a Northward configuration of the IMF, identifying the region of the BS (color scale of the density distribution), field lines of the Hermean magnetic field (red lines), IMF (pink lines) and solar wind stream lines (green lines). The arrows indicate the orientation of the Hermean and interplanetary magnetic fields and the dashed white line the beginning of the simulation domain.

\begin{figure}[h]
\centering
\includegraphics[width=\columnwidth]{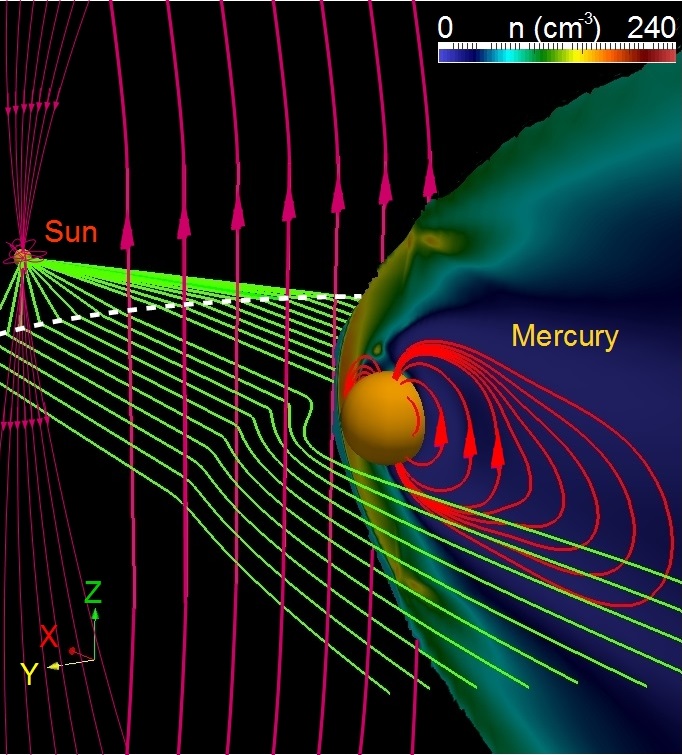}
\caption{3D view of the system. Density distribution (color scale), field lines of the Hermean magnetic field (red lines), IMF (pink lines) and solar wind stream lines (green lines). The arrows indicate the orientation of the Hermean and interplanetary magnetic fields (case Bz). Dashed white line shows the beginning of the simulation domain.}
\end{figure}

\begin{table}[h]
\centering
\begin{tabular}{c c c c c c c c}
$n$ (cm$^{-3}$) & $T$ (K) & $\mathrm{v}$ (km/s) & $M_{s}$  \\ \hline
$60$ & $58000$ & $250$ & $6.25$ \\
\end{tabular}
\caption{Hydrodynamic parameters of the SW}
\end{table}

In the following we identify the Mercury-Sun orientation as Bx simulations, the Sun-Mercury orientation as Bxneg simulations, the Northward orientation respect to Mercury's magnetic axis as Bz simulations (shown in figure 1 example), the Southward orientation as Bzneg simulations, the orientation perpendicular to previous two cases on the planet orbital plane as By (East) and Byneg (West) simulations. The IMF intensity of the model is denoted by a number attached to the orientation label, for example the simulation Bz3 identifies a Northward orientation of the IMF with module 30 nT. All the simulations and parameters are summarized in appendix A. We include as reference case a simulation without IMF (B0).

\begin{figure}[t!]
\centering
\includegraphics[width=0.45\textwidth]{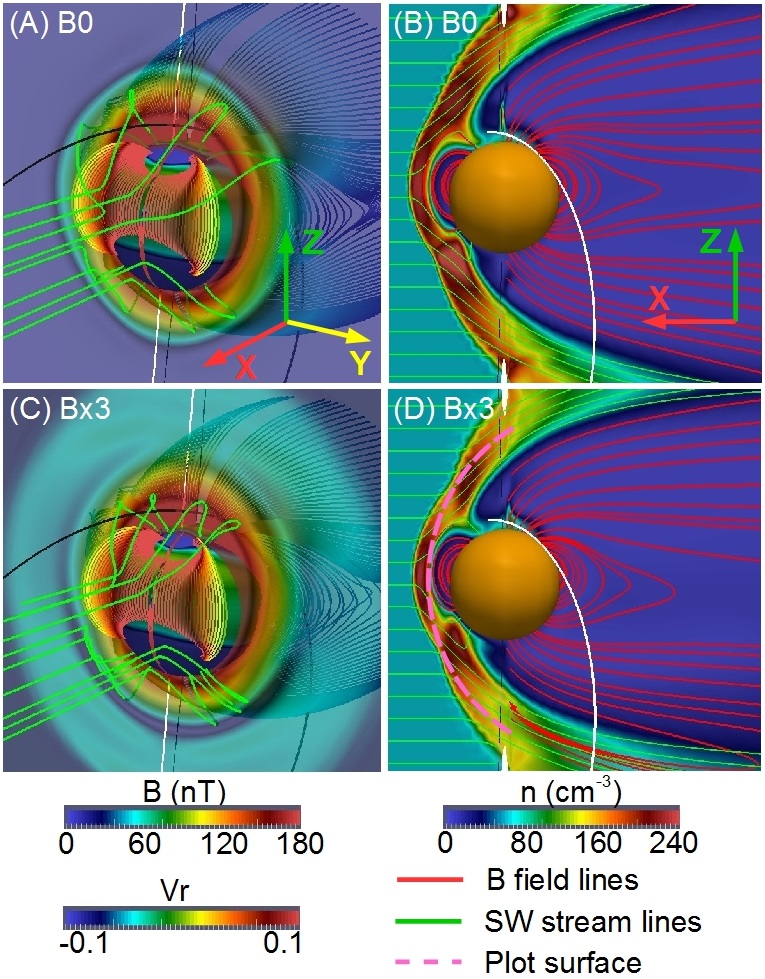}
\caption{Hermean magnetic field lines with the intensity imprinted on the field lines by a color scale for the reference case (A) and simulation Bx3 (C). Magnetic field intensity at the frontal plane $X = 0.3R_{M}$. SW stream lines (green). Inflow/outflow regions on the planet surface (blue/red). Polar plot of the density distribution (displaced $0.1R_{M}$ in $Y$ direction) for the reference case (B) and simulation Bx3 (D). Dashed pink curve indicates the surface plotted in figures 3 and 4.}
\end{figure}

In Figure 2 we illustrate the interaction of the IMF and magnetospheric field for the reference case and Bx3 simulation. Panels A and C show the Hermean magnetic field lines (magnetic field lines are color-coded with the magnetic field amplitude), SW stream lines (green lines), magnetic field intensity at the frontal plane $X = 0.3R_{M}$ and inflow/outflow (blue/red) regions on the planet surface. Reconnection regions are identified as blue colors at the frontal plane, not observed in the reference case and located in the South of the magnetosphere for the Bx3 simulation. Panels B and D show polar plots of the density distribution including the magnetic field lines of the planet (red lines) and SW stream lines (green lines), identifying the regions of the bow shock, magnetosheath and magnetopause. The dashed pink line indicates the surface plotted in figures 3 and 4 for the Bx3 simulation.

\begin{figure*}[t!]
\centering
\includegraphics[width=0.8\textwidth]{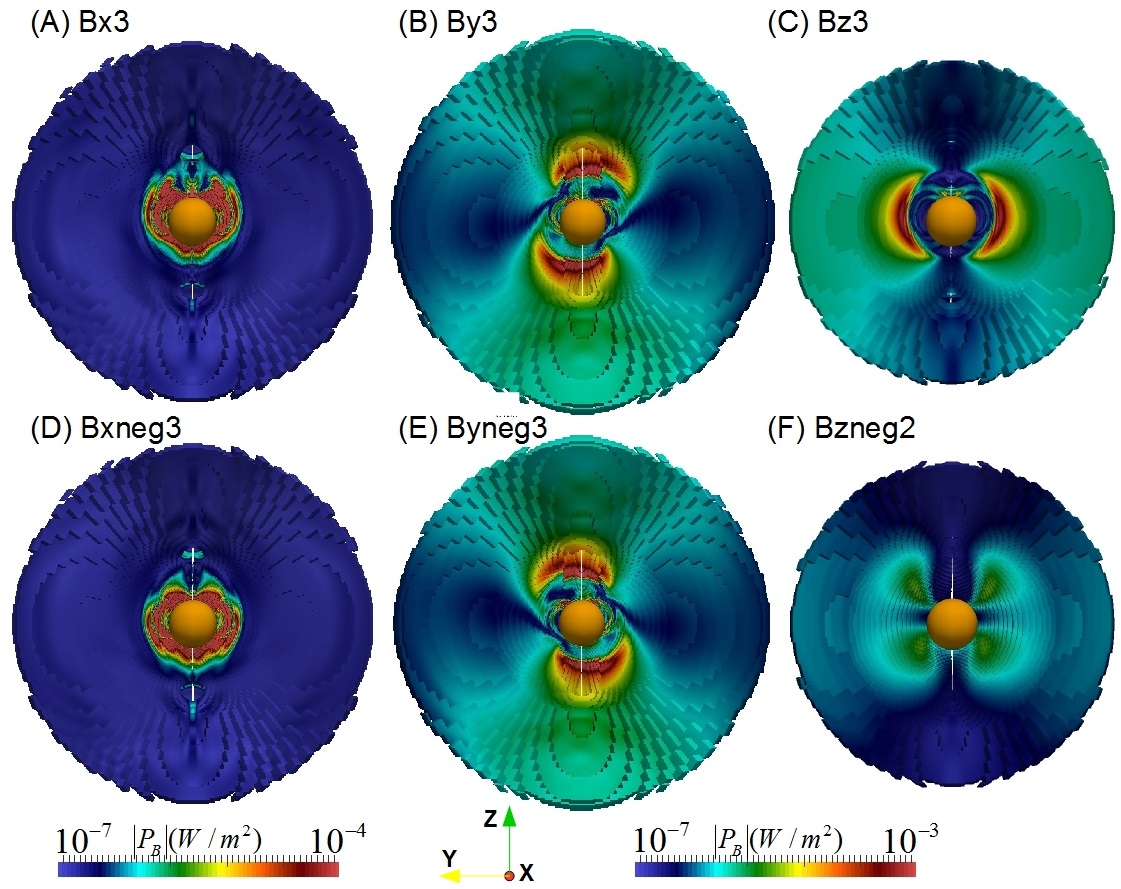}
\caption{Frontal view of the magnetic energy flux on the planet day side for different IMF orientations. We observe the magnetic energy flux distribution towards the planet-Sun direction from the night side. 1st color bar is related to (A) and (D) panels, 2nd color bar to the other 4 cases. The plotted surface is defined between the bow shock and the magnetopause where the magnetic energy flux reaches its maxima (see fig. 2, panels B and D, pink dashed line).}
\end{figure*}

Figure 3 shows a frontal view of the magnetic energy flux on the planet day side ($P_{B}(DS)$) for different IMF orientations, observed towards the planet-Sun direction from the planet night side. There is a strong dependency of $P_{B}(DS)$ hot spot distribution with the IMF orientation. Local minima of the magnetic energy flux are correlated with a local decrease of the Hermean magnetic field intensity, although local maxima are correlated with a local enhancement of the Hermean magnetic field. For Bx3 (panel A) and Bxneg3 (panel D) IMF orientations, $P_{B}(DS)$ hot stops are located close to the planet. There is a North-South asymmetry due to the presence of a reconnection region, located in the South (North) of the magnetosphere for the Bx3 (Bxneg3) case, identified as a local decrease of $P_{B}(DS)$. Bx3 and Bxneg3 orientation weakly affect the Hermean magnetosphere topology at low-middle latitudes, only nearby the poles, reason why both cases appear to be similar. Compared with other IMF orientations, the local maximum of $P_{B}(DS)$ is almost one order of magnitude smaller. Models with By3 (panel B) and Byneg3 (panel E) IMF orientations have $P_{B}(DS)$ hot spots nearby the poles, showing the same East-West asymmetry than the magnetosphere, with the reconnection regions located in the planet sides (local minimum of $P_{B}(DS)$). Bz3 (panel C) and Bzneg2 (panel F) cases show different distributions of the $P_{B}(DS)$ hot spots, along the planet sides for Bz3 orientation (reconnection regions located near the poles) and a quadripolar structure in the Bzneg2 model (reconnection region at the planet equator). These results indicate a large variability of the radio emission distribution on the planet dayside with the IMF orientation. Thus, future radio emission observation from exoplanets must show a strong dependency with the host star magnetic activity.

\begin{figure*}[t!]
\centering
\includegraphics[width=0.8\textwidth]{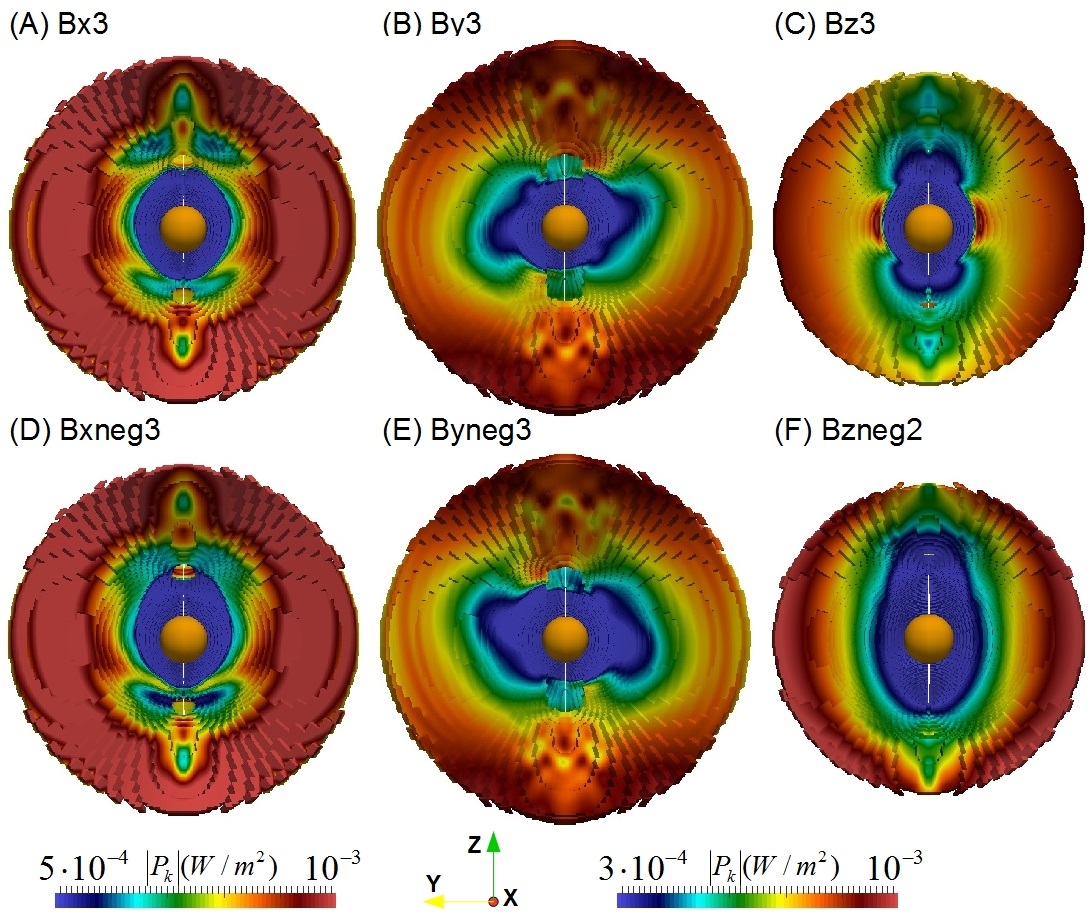}
\caption{Frontal view of the kinetic energy flux on the planet day side for different IMF orientations. We observe the kinetic energy flux distribution towards the planet-Sun direction from the night side.Color scale on the left for figures (A) and (D). The plotted surface is defined between the bow shock and the magnetopause where the kinetic energy flux reaches its maxima (see fig. 2, panels B and D, pink dashed line).}
\end{figure*}

Figure 4 shows a frontal view of the kinetic energy flux on the planet day side ($P_{k}(DS)$) for different IMF orientations, observed towards the planet-Sun direction from the planet night side. There is a local enhancement (decrease) of $P_{k}(DS)$ associated with a local decrease (enhancement) of the magnetospheric field. $P_{k}(DS)$ is linked with the dynamic pressure of the SW although the local enhancements observed for all the IMF orientations is determined by the magnetosphere topology and IMF orientation. This statement is valid in case of exoplanets as long as we consider a purely radial outflow of stellar wind between the star and the exoplanet, condition that could not be fulfilled for exoplanet with eccentric orbits.

\begin{figure*}[t!]
\centering
\includegraphics[width=0.8\textwidth]{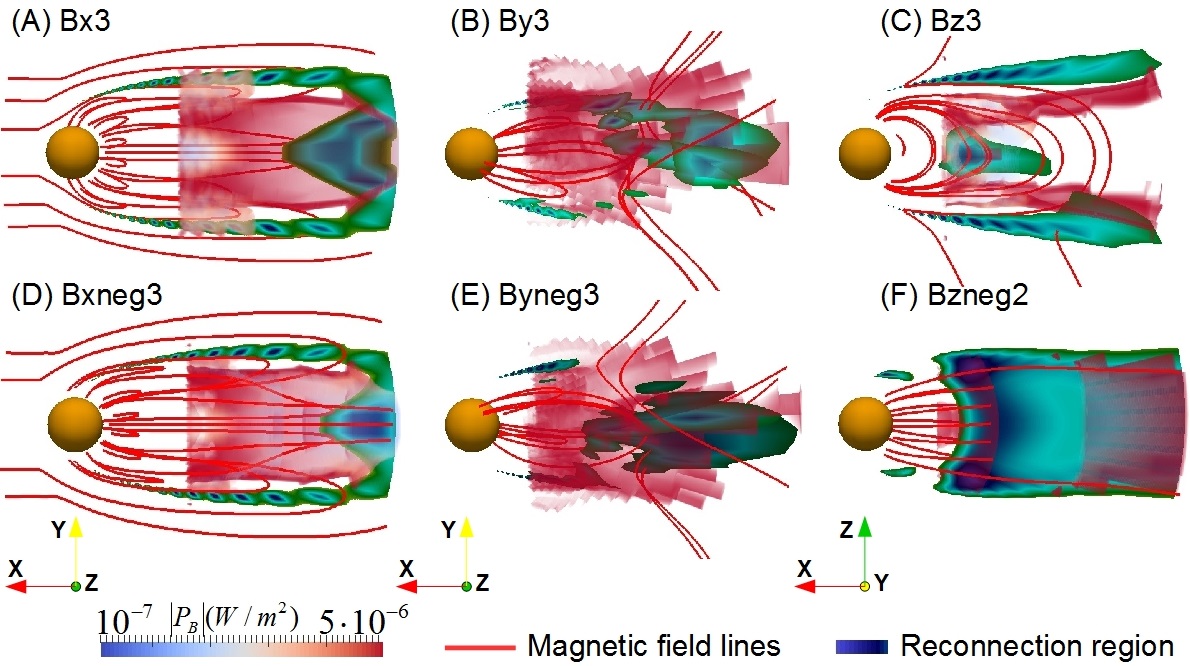}
\caption{Magnetic energy flux on the planet night side ($P_{B}(NS)$). Blue color indicates the reconnection region (iso-surface with magnetic field values between $0 - 7$ nT). Magnetic field lines of the Hermean and IMF in red.}
\end{figure*}

Figure 5 shows the magnetic energy flux ($P_{B}(NS)$) on the planet night side. $P_{B}(NS)$ distribution shows a local maxima between the reconnection regions of the magnetotail and magnetopause for all IMF orientations. If the magnetotail is slender and stretched, case of By3 (panel B), Byneg3 (panel E) and Bzneg2 (panel F) IMF orientations, the magnetic tension is larger leading to a more efficient dissipation of the magnetic energy and larger $P_{B}(NS)$ values. We don't include the analysis of the kinetic energy flux ($P_{k}(NS)$), whose local maxima is located in the magnetotail reconnection region, because the expected radio emission associated with $[P_{k}]$ is small compared to $[P_{B}]$ (see table 4). Consequently, we must presume greater radio emission \textcolor{red}{on} the exoplanets night side if the IMF orientation drives slender and stretched magnetotail topologies. At the same time, for exoplanets with more intense magnetic field hosted by stars with stronger magnetic activity than the Sun, radio emission \textcolor{red}{on} planet night side should dominate because magnetic and kinetic energy fluxes in the magnetotail reconnection region are enhanced.

Table 3 shows the magnetic and kinetic net power dissipated on the planet day side and the reconnection region of the magnetotail on the planet night side. The net magnetic energy is negative because the IMF interaction with the Hermean magnetosphere leads in all models to a net erosion of the magnetospheric field, so the system loses locally magnetic energy. The net kinetic energy is negative too on planet day side because the solar wind is decelerated as soon as it reaches the Hermean magnetosphere, losing kinetic energy. The opposite scenario is observed in the magnetotail X point, there is a net gain of kinetic energy by the plasma because it is accelerated in the reconnection region. Bz3 orientation is the only case with $[P_{k}(NS)] < 0$, because it is the configuration with the magnetotail X point located the furthest from the planet, leading to a smaller magnetic energy flux and plasma acceleration in the reconnection region. $[P_{B}(DS)]$ and $[P_{k}(DS)]$ are larger than $[P_{B}(NS)]$ and $[P_{k}(NS)]$ for all IMF orientations except Bx3 case. $[P_{B}(DS)]$ for Bx3 and Bxneg3 IMF orientations is one order of magnitude smaller than in other models. $[P_{k}]$ is larger than $[P_{B}]$ on the planet day side although $[P_{B}]$ is larger than $[P_{k}]$ in the reconnection region of the magnetotail on the night side for the models By3, Byneg3 and Bzneg2. The IMF orientation that drives the highest net power dissipation is Bzneg2 configuration, while Bx3 orientation leads to the smallest. 

\begin{table*}[t!]
\centering
\begin{tabular}{c | c c | c c }
Model & $[P_{k}(DS)]$ ($10^{8}$ W) & $[P_{B}(DS)]$ ($10^{8}$ W) & $[P_{k}(NS)]$ ($10^{7}$ W) &  $[P_{B}(NS)]$ ($10^{7}$ W) \\ \hline
Bx3 & $-4.60$ & $-0.13$ & $3.11$ & $-1.61$ \\
Bxneg3 & $-4.71$ & $-0.25$ & $3.48$ & $-0.84$ \\
By3 & $-5.93$ & $-2.81$ & $3.30$ & $-7.95$ \\
Byneg3 & $-5.94$ & $-2.83$ & $3.50$ & $-8.05$ \\
Bz3 & $-5.91$ & $-2.95$ & $-6.19$ & $-4.31$ \\
Bzneg2 & $-11.4$ & $-2.94$ & $7.40$ & $-13.5$ \\
\end{tabular}
\caption{Net magnetic and kinetic power dissipated on the planet day side and the reconnection region at planet night side. SW hydrodynamic parameters: $\rho = 60$ cm$^{-3}$, $v = 250$ km/s and $T = 58000$ K. Negative (positive) sign indicates a net lost (gain) of energy by the plasma.}
\end{table*}

In table 4 we show the expect radio emission calculated from the net magnetic and kinetic power dissipated on the planet day side and the reconnection region at planet night side using radiometric Bode's law \citep{Zarka3,Farrell}:

\begin{equation} \label{eq:10}
[P(DS)] = a [P_{k}(DS)] + b [P_{B}(DS)]
\end{equation}
\begin{equation} \label{eq:11} 
[P(NS)] = a [P_{k}(NS)] + b [P_{B}(NS)]
\end{equation} 
with $a$ and $b$ efficiency ratios assuming a linear dependency of $[P_{k}]$ and $[P_{B}]$ with $[P]$. Observations of other planets in the solar system are explained by two possible combination of parameters ($a = 10^{-5}$, $b=0$) or ($a = 0$, $b=2\cdot10^{-3}$) \citep{Zarka4}.

\begin{table*}[t!]
\centering
\begin{tabular}{c | c c | c c}
Model & $[P(NS)]$ ($10^{5}$ W) & $[P(DS)]$ ($10^{5}$ W) & $[P(NS)]$ ($10^{2}$ W) & $[P(DS)]$ ($10^{3}$ W) \\ \hline
Bx3 & $0.32$ & $0.28$ & 3.11 & 4.60 \\
Bxneg3 & $0.17$ & $0.48$ & 3.48 & 4.71 \\
By3 & $1.59$ & $5.64$ & 3.30 & 5.93 \\
Byneg3 & $1.61$ & $5.68$ & 3.50 & 5.94 \\
Bz3 & $0.86$ & $5.92$ & 6.19 & 5.91 \\
Bzneg2 & $2.70$ & $5.88$ & 7.40 & 11.4 \\
\end{tabular}
\caption{Expected radio emission on the planet day side and in the reconnection region of the magnetotail on the planet night side for the efficiency ratios ($a = 0$, $b=2\cdot10^{-3}$, second and third columns, and ($a = 1\cdot10^{-5}$, $b=0$), fourth and fifth columns.}
\end{table*}

$[P_{B}]$ is similar to $[P_{k}]$ for all IMF orientations (except Bx3 and Bxneg3 cases with $[P_{k}]$ one order of magnitude larger than $[P_{B}]$), so the combination of efficiency ratios ($a = 10^{-5}$, $b=0$) is several orders of magnitude smaller than ($a = 0$, $b=2\cdot10^{-3}$), see table 4. $[P(NS)]$ for ($a = 0$, $b=2\cdot10^{-3}$) combination is almost 5 times smaller than $[P(DS)]$, pointing out that the main source of radio emission in the Hermean magnetosphere should be the strong compression of the planet magnetic field lines by the SW on the planet dayside, as well as local amplifications of the Hermean magnetic field by the IMF. To determine the combination of efficiency ratios that better fits radio emission from Mercury, it is required to perform in situ measurements comparing radio emission on the planet day and night side. If the radio emission on the planet day side is in average 10 times larger than the radio emission at the night side, ($a = 0$, $b = 2 \cdot 10^-{3}$) combination is the best description, although if the average difference is closer to 100 times, ($a = 10^-{5}$, $b = 0$) combination is the best approach. The ratio between radio emission on the planet day and night side (DS-NS ratio) must change in exoplanets with more intense magnetic fields, hosted by stars with stronger magnetic activity than the Sun. It is mandatory to study the ratio of DS-NS radio emission in giant planets of the solar system and determine which combination of efficiency ratios fits better the observations.

\subsection{Intensity of the interplanetary magnetic field}

We analyze the influence of the IMF amplitude variation on radio emission. We perform 13 new simulation using the same SW hydro parameters and IMF orientations than in the previous section (see table 2), varying the IMF intensity from $10$ to $30$ nT.

Figure 6 shows $[P(DS)]$ (panel A) and $[P(NS)]$ (panel B) ($a = 0$ and $b=2\cdot10^{-3}$ combination) for models with different IMF orientations and modules from $10$ to $30$ nT. $[P(DS)]$ and $[P(NS)]$ increase with the IMF module for all orientations except Bx and Bxneg cases, almost insensitive to the IMF intensity variation. The radio emission enhancement is linked with a local increase of the magnetospheric field by the IMF, amplification not observed in Bx and Bxneg simulations, because in these configurations the Herman magnetic field and IMF are perpendicular at low and middle latitudes, with the reconnection region located close to the poles. Bz orientation shows a decrease of $[P(NS)]$ in the simulations with IMF module $10$ and $20$ nT, because the magnetotail is wider and less stretched, leading to a decrease of the magnetic field lines tension. The tendency is inverted in Bz3 simulation due to the induced North-South stretching of the magnetotail.

 \begin{figure*}[t!]
\centering
\includegraphics[width=0.6\textwidth]{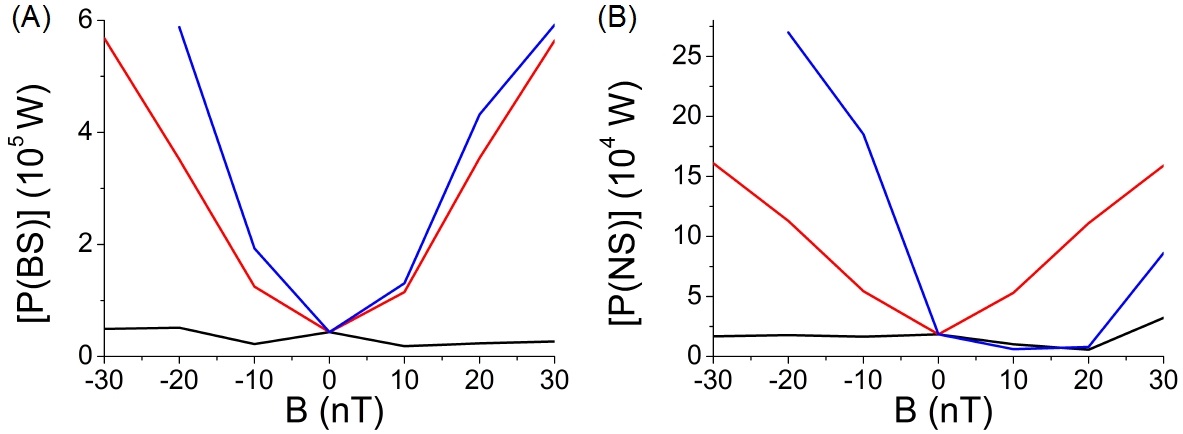}
\caption{Radio emission on the planet day side (A) and in reconnection region of the magnetotail on the planet night side (B) with radio efficiency parameters ($a = 0$, $b=2\cdot10^{-3}$) for models with different IMF orientation and modules from $10$ to $30$ nT. The reference case without IMF is included. Bx orientations (black line), By orientations (red line) and Bz orientation (blue line).}
\end{figure*}

\section{Effect of the hydro parameters of the solar wind}

This section studies the effect of the SW hydrodynamic parameters on radio emission generation. We perform 6 new simulations fixing IMF module and orientation, $\vec{B}_{SW} = (4,1,6)$ nT. We choose a IMF configuration with weak module ($7.28$ nT) to maximize the effect of the SW hydro parameters, and a mixed orientation to show a more realistic case. SW parameters of reference case are the same than in Table 3. The name of the model is given by the hydro parameter that is modified respect to reference case (see Table 5).

\begin{table*}[t!]
\centering
\begin{tabular}{c | c |c c | c | c}
Model & Mod. Parameter & $[P_{k}(DS)]$ ($10^{8}$ W) & $[P_{B}(DS)]$ ($10^{8}$ W) & $[P]$ ($10^{6}$ W) & $[P]$ ($10^{3}$ W) \\ \hline
$\rho_{min}$ & $n = 12$ cm$^{-3}$ & -1.03 & -0.50 & 0.10 & 1.03 \\
$\rho_{max}$ & $n = 180$ cm$^{-3}$ & -8.24 & -10.30 & 2.06 & 8.24 \\
$\mathrm{v}_{min}$ & $\mathrm{v} = 200$ km/s & -2.42 & -0.72 & 0.14 & 2.42 \\
$\mathrm{v}_{max}$ & $\mathrm{v} = 500$ km/s & -18.1 & -9.03 & 1.80 & 18.1 \\
$T_{min}$ & $T = 2 \cdot 10^{4}$ K & -3.66 & -0.68 & 0.14 & 3.66 \\
$T_{max}$ & $T = 18 \cdot 10^{4}$ K & -4.78 & -0.98 & 0.20 & 4.78 \\
\end{tabular}
\caption{Second column indicates the hydrodynamic parameter modified respect to reference case ($n = 60$ cm$^{-3}$, $\mathrm{v} = 250$ km/s and $T = 58000$ K). Net magnetic and kinetic power dissipated on the planet day side (second and third columns). Expected radio emission on the planet day side with radio efficiency parameters ($a = 0$, $b=2\cdot10^{-3}$), fourth column, and ($a = 10^{-5}$, $b = 0$), fifth column.}
\end{table*}

Figure 7 shows a frontal view of the magnetic energy flux on the planet dayside for different configurations of SW hydro parameters, observed towards the planet-Sun direction from the night side. 

\begin{figure*}[t!]
\centering
\includegraphics[width=0.8\textwidth]{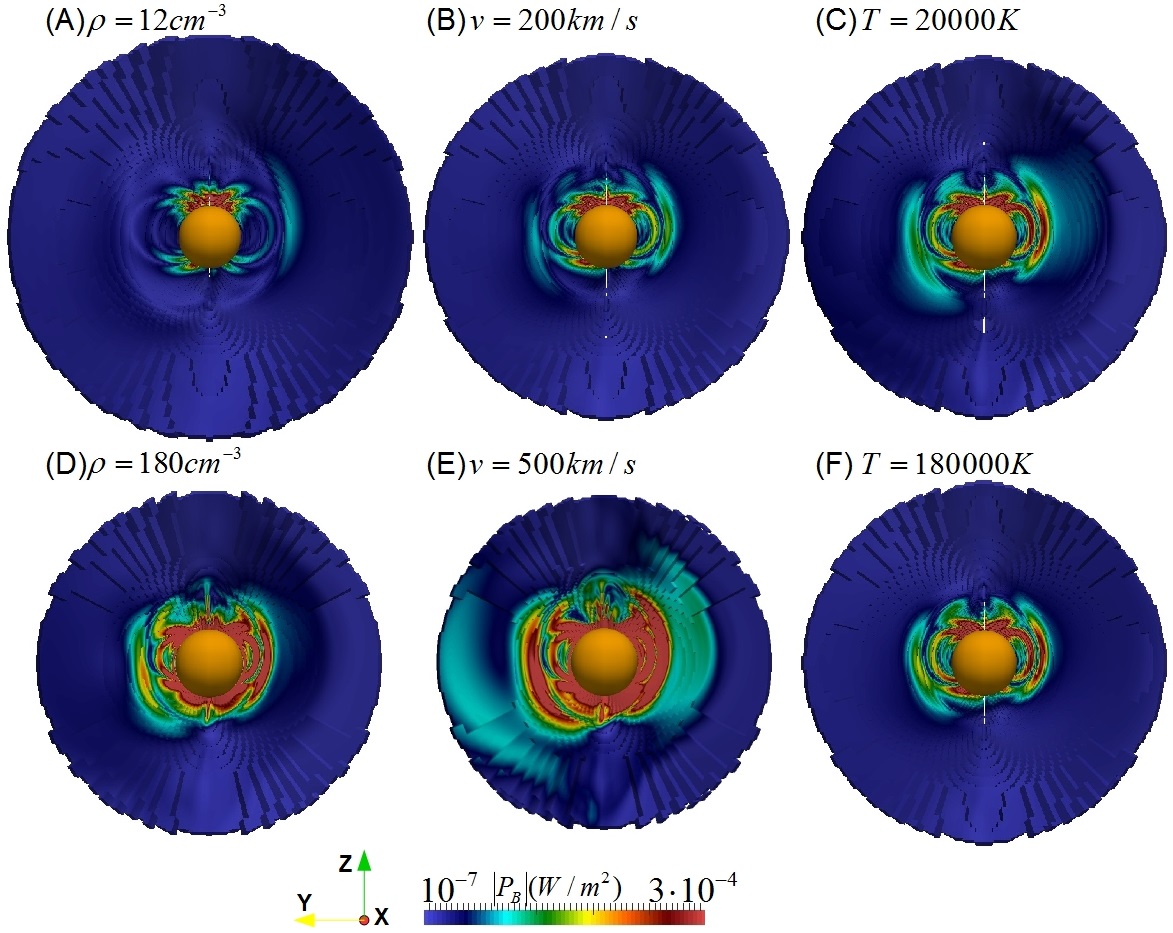}
\caption{Frontal view of the magnetic energy flux on the planet day side for simulations with different SW hydro parameters. We observe the magnetic energy flux distribution towards the planet-Sun direction from the night side. The plotted surface is defined between the bow shock and the magnetopause where the magnetic energy flux reaches its maxima (see fig. 2, panels B and D, pink dashed line).}
\end{figure*}

$P_{B}(DS)$ distribution is similar for all simulations, pointing out that the location of radio emission hot spots is dictated by the IMF orientation and not by the SW hydro parameters. SW hydro parameters can enhance or weaken the local maxima but the distribution is only slightly modified. $P_{B}(DS)$ local maxima are enhanced if the SW dynamic pressure ($q = \rho \mathrm{v}^2/2$) is larger, as in the simulations $\rho_{max}$ and $\mathrm{v}_{max}$, compared to low SW dynamic pressure configurations as $\rho_{min}$ and $\mathrm{v}_{min}$. $P_{B}(DS)$ local maxima are almost the same for $T_{max}$ and $T_{min}$ simulations, because the temperature doesn't affect the Hermean magnetic field compression, only leads to a decompression of the magnetosheath, increasing the SW sound velocity and reducing the sonic Mach number. In summary, if we follow the evolution of the hot spot distribution as the SW dynamic pressure increases (panels A, B, C, F, D and E in this order), hot spot intensity increases but the distribution remains almost the same.

Table 5 shows $[P_{B}(DS)]$, $[P_{k}(DS)]$ and $[P(DS)]$ with radio efficiency parameters ($a = 0$, $b=2\cdot10^{-3}$) and ($a = 10^{-5}$, $b = 0$) for simulations with different SW hydro parameters. ($a = 0$, $b=2\cdot10^{-3}$) combination leads to a radio emission two orders of magnitude larger than ($a = 10^{-5}$, $b = 0$). The configurations with the largest $[P(DS)]$ are $\rho_{max}$ and $\mathrm{v}_{max}$, more than one order of magnitude larger than configurations with lower dynamic pressure as $\rho_{min}$ and $\mathrm{v}_{min}$. $T_{min}$ and $T_{max}$ simulations show similar $[P(DS)]$ values, slightly larger in the simulation with less compressed magnetosheath. $[P_{k}(DS)]$ is 2 to 5 times larger than $[P_{B}(DS)]$ for all cases except $\rho_{max}$ case. These results demonstrate that radio emissions are sensitive to the time variation of the SW hydro parameters.

\section{Conclusions and discussion}
\label{Conclusions}

The analysis of the power dissipation in 3D simulations of the solar wind interaction with the Hermean magnetosphere shows that hot spot distribution of the radio emission is determined by the IMF orientation, although the intensity of the local maxima is dictated by the IMF intensity and SW hydro parameters. The orientation of the IMF establishes the location of the reconnection regions in the Hermean magnetosphere. A local enhancement of the magnetospheric field on the day side is correlated with a maxima of the magnetic energy flux and a minima of the kinetic energy flux.

The net dissipated power on the planet day side is larger than in the reconnection region of the magnetotail on the planet night side, by almost one order of magnitude larger. This proportion can be very different in other planets of the solar system with more intense magnetic field, facing a SW of lower dynamic pressure, as in the case of the Earth of giant gaseous planets, leading to configurations with stronger radio emission from the reconnection region of the magnetotail and weaker from the planet dayside. 

Radio emission on the planet day side is maximized for any IMF orientation different than Sun-Mercury or Mercury-Sun directions. These two cases show a radio emission more than one order of magnitude smaller. The magnetic field lines of the IMF and Hermean magnetosphere are almost perpendicular at low-middle latitudes, avoiding an amplification of the magnetospheric field. This result is consistent with the weak dependency of radio emission and IMF module for these orientations. The other IMF orientations lead to local enhancements of the magnetospheric field on the planet dayside, correlated with hot spots of radio emission.

The net dissipated power is dominated by the magnetic component, pointing out that on the planet night side the main source of radio emission is the magnetic tension of the magnetotail field lines. The consequence is a stronger radio emission in such magnetosphere configurations with stretched magnetotail and reconnection X point located close to the planet, see cases with Bzneg, By and Byneg IMF orientations.

The combination of efficiency ratios that lead to larger radio emission is ($a = 0$, $b=2\cdot10^{-3}$), several order of magnitude larger than the expected radio emission for the combination ($a = 10^{-5}$, $b=0$). If the radio emission on the planet night side is around $10 \%$ of total radio emission, ($a = 0$, $b = 2 \cdot 10^-{3}$) combination describes better the conversion of dissipated power on radio emission, but if it is less than $1 \%$ the correct formulation must contain the combination ($a = 10^-{5}$, $b = 0$). In situ observations are required to confirm the correct efficiency ratios combination.

The results obtained are compatible with the predicted values by \citet{Zarka3}, expecting a radio emission around $10^{6}$ W. This is the case for all the simulations expect configurations with the IMF oriented in Sun-Mercury of Mercury-Sun directions, or models with low dynamic pressure and weak IMF intensity. Simulations with large dynamic pressure and IMF intensities larger than $10$ nT not oriented in Sun-Mercury of Mercury-Sun directions, show radio emissions from $1$ to $2 \cdot 10^{6}$ W. 

Radiometric Bode's law for Mercury's magnetosphere considering a stand off distance of $1.5R_{M}$ predicts a net kinetic energy dissipation of $6 \cdot 10^{10}$ W and a net magnetic energy dissipation of $4-40 \cdot 10^{8}$ W (depending on the IMF intensity). The net kinetic energy dissipation calculated from the simulations is smaller than the expected value by Bode's law, consequence of the complex flows reproduced in the simulations with regions of plasma deceleration on the day side and acceleration in the magnetotail reconnection region. On the other side, the simulations and Bode's law lead to similar results for the net magnetic energy dissipation, pointing out that the model is able to provide a reasonable guess of the magnetic energy dissipation on the day side and magnetotail reconnection regions.

A systematic study of the conversion efficiency between dissipated power and radio emission in the planets of the solar system is a key study to calibrate the radio emission from exoplanets. It is expected that radio emission data brings constrains on the intensity and topology of exoplanets magnetic fields, information required to study the potential habitability of exoplanets, directly linked with the presence of permanent and strong enough magnetic fields to shield the planet surface and atmosphere from the stellar wind erosion. We analyze the effect of the IMF orientation and module, as well as SW hydro parameters on radio emission generation, showing the large variability induced by these factors on hot spot distribution and intensity. A generalization of present study to the specific environment of the exoplanets, stellar wind of host star and expected IMF intensity and orientation at exoplanet orbit, is mandatory to reconstruct the topology and intensity of the intrinsic magnetic field from radio emission data when available \citep{Hess,Zarka6}. 

\begin{acknowledgements}
The research leading to these results has received funding from the European Commission's Seventh Framework Programme (FP7/2007-2013) under the grant agreement SHOCK (project number 284515), the grant agreement SPACEINN (project number 312844) and ERC STARS2 (207430). We extend our thanks to CNES for Solar Orbiter and PLATO science support as well as INSU/PNST for our grant. We acknowledge GENCI allocation 1623 for access to supercomputer where most simulations were run and DIM-ACAV for supporting ANAIS project and our graphics/post-analysis and storage servers.
\end{acknowledgements}

\begin{appendix}
Summary of simulations parameters:

\begin{table}[h]
\centering
\begin{tabular}{c | c c c c}
Model & $\vec{B}$ (nT) & $n$ (cm$^{-3}$) & $T$ ($10^{5}$ K) & $\mathrm{v}$ (km/s) \\ \hline
Reference & (0, 0, 0) & $60$ & 0.58 & 250 \\
Bx & (10, 0, 0) & $60$ & 0.58 & 250 \\
Bx2 & (20, 0, 0) & $60$ & 0.58 & 250 \\
Bx3 & (30, 0, 0) & $60$ & 0.58 & 250 \\
Bxneg & (-10, 0, 0) & $60$ & 0.58 & 250 \\
Bxneg2 & (-20, 0, 0) & $60$ & 0.58 & 250 \\
Bxneg3 & (-30, 0, 0) & $60$ & 0.58 & 250 \\
By & (0, 10, 0) & $60$ & 0.58 & 250 \\
By2 & (0, 20, 0) & $60$ & 0.58 & 250 \\
By3 & (0, 30, 0) & $60$ & 0.58 & 250 \\
Byneg & (0, -10, 0) & $60$ & 0.58 & 250 \\
Byneg2 & (0, -20, 0) & $60$ & 0.58 & 250 \\
Byneg3 & (0, -30, 0) & $60$ & 0.58 & 250 \\
Bz & (0, 0, 10) & $60$ & 0.58 & 250 \\
Bz2 & (0, 0, 20) & $60$ & 0.58 & 250 \\
Bz3 & (0, 0, 30) & $60$ & 0.58 & 250 \\
Bzneg & (0, 0, -10) & $60$ & 0.58 & 250 \\
Bzneg2 & (0, 0, -20) & $60$ & 0.58 & 250 \\
$\rho_{min}$ & (4, 1, 6) & $12$ & 0.58 & 250 \\
$\rho_{max}$ & (4, 1, 6) & $180$ & 0.58 & 250 \\
$\mathrm{v}_{min}$ & (4, 1, 6) & $60$ & 0.58 & 200 \\
$\mathrm{v}_{max}$ & (4, 1, 6) & $60$ & 0.58 & 500 \\
$T_{min}$ & (4, 1, 6) & $60$ & 0.20 & 250 \\
$T_{max}$ & (4, 1, 6) & $60$ & 1.80 & 250 \\

\end{tabular}
\caption{Summary of simulations parameters.}
\end{table}
 
Bzneg3 simulation is not included in the text because the magnetopause is located on the planet surface. Present model lacks the Physics required to correctly reproduce that scenario.
\end{appendix}

%\newpage
%\bibliographystyle{aa}
%\bibliography{/Publications/Mercury/Radio emission/A&A/aa}

\bibliographystyle{aa}
\bibliography{References}

\begin{thebibliography}{38}
\expandafter\ifx\csname natexlab\endcsname\relax\def\natexlab#1{#1}\fi

\bibitem[{{Anderson}(2008{\natexlab{a}})}]{Anderson2}
{Anderson}, B.~J. 2008{\natexlab{a}}, J. Geophys. Res., 117, E00L12

\bibitem[{{Anderson}(2008{\natexlab{b}})}]{Anderson3}
{Anderson}, B.~J. 2008{\natexlab{b}}, Science, 321, 82

\bibitem[{{Anderson}(2011)}]{Anderson}
{Anderson}, B.~J. 2011, Science, 333, 1859

\bibitem[{{Anderson} {et~al.}(2012){Anderson}, {Johnson}, {Korth}, {Winslow},
  {Borovsky}, {Purucker}, {Slavin}, {Solomon}, {Zuber}, \&
  {McNutt}}]{Anderson4}
{Anderson}, B.~J., {Johnson}, C.~L., {Korth}, H., {et~al.} 2012, Journal of
  Geophysical Research (Planets), 117, E00L12

\bibitem[{{Baker}(2009)}]{Baker}
{Baker}, D.~N. 2009, J. Geophys. Res., 114, A10101

\bibitem[{{Baker}(2011)}]{Baker2}
{Baker}, D.~N. 2011, Planet. Space Sci., 59, 2066

\bibitem[{{Dedner}(2002)}]{Dedner}
{Dedner}, A. 2002, J. Comput. Phys., 175, 645

\bibitem[{{Farrell}(1999)}]{Farrell}
{Farrell}, W.~M. 1999, Geophys. Res. Lett., 104, 14025

\bibitem[{{Fujimoto}(2007)}]{Fujimoto}
{Fujimoto}, M. 2007, Space Sci. Rev., 132, 529

\bibitem[{{Hess}(2011)}]{Hess}
{Hess}, S. L.~G. 2011, A \& A, 531, A29

\bibitem[{{Jia} {et~al.}(2015){Jia}, {Slavin}, {Gombosi}, {Daldorff}, {Toth},
  \& {Holst}}]{2015JGRA..120.4763J}
{Jia}, X., {Slavin}, J.~A., {Gombosi}, T.~I., {et~al.} 2015, Journal of
  Geophysical Research (Space Physics), 120, 4763

\bibitem[{{Johnson}(2012)}]{Johnson}
{Johnson}, C.~L. 2012, J. Geophys. Res., 117, 2076

\bibitem[{{Kabin} {et~al.}(2000){Kabin}, {Gombosi}, {DeZeeuw}, \&
  {Powell}}]{2000Icar..143..397K}
{Kabin}, K., {Gombosi}, T.~I., {DeZeeuw}, D.~L., \& {Powell}, K.~G. 2000,
  Icarus, 143, 397

\bibitem[{{Kabin} {et~al.}(2008){Kabin}, {Heimpel}, {Rankin}, {Aurnou},
  {G{\'o}mez-P{\'e}rez}, {Paral}, {Gombosi}, {Zurbuchen}, {Koehn}, \&
  {DeZeeuw}}]{2008Icar..195....1K}
{Kabin}, K., {Heimpel}, M.~H., {Rankin}, R., {et~al.} 2008, Icarus, 195, 1

\bibitem[{{Kidder} {et~al.}(2008){Kidder}, {Winglee}, \&
  {Harnett}}]{2008JGRA..113.9223K}
{Kidder}, A., {Winglee}, R.~M., \& {Harnett}, E.~M. 2008, Journal of
  Geophysical Research (Space Physics), 113, A09223

\bibitem[{{Mignone}(2007)}]{Mignone}
{Mignone}, A. 2007, ApJS, 170, 228

\bibitem[{{M\"{u}ller} {et~al.}(2011){M\"{u}ller}, {Simon}, {Motschmann},
  {Schüle}, {Glassmeier}, \& {Gringle}}]{Muller2011946}
{M\"{u}ller}, J., {Simon}, S., {Motschmann}, U., {et~al.} 2011, Computer
  Physics Communications, 182, 946

\bibitem[{{M\"{u}ller} {et~al.}(2012){M\"{u}ller}, {Simon}, {Wang},
  {Motschmann}, {Heyner}, {Schüle}, {Ip}, {Kleindienst}, \&
  {Gringle}}]{Muller2012666}
{M\"{u}ller}, J., {Simon}, S., {Wang}, Y.~C., {et~al.} 2012, Icarus, 218, 666

\bibitem[{{Odstrcil}(2003)}]{Odstrcil}
{Odstrcil}, D. 2003, Adv. Space Res., 32, 497

\bibitem[{{Parker}(1958)}]{Parker}
{Parker}, E.~N. 1958, Astrophys. J., 128, 664

\bibitem[{{Pizzo}(2011)}]{Pizzo}
{Pizzo}, V. 2011, Space Weather, 9, S03004

\bibitem[{{Richer} {et~al.}(2012){Richer}, {Modolo}, {Chanteur}, {Hess}, \&
  {Leblanc}}]{2012JGRA..11710228R}
{Richer}, E., {Modolo}, R., {Chanteur}, G.~M., {Hess}, S., \& {Leblanc}, F.
  2012, Journal of Geophysical Research (Space Physics), 117, A10228

\bibitem[{{Saur} {et~al.}(2013){Saur}, {Grambusch}, {Duling}, {Neubauer}, \&
  {Simon}}]{Saur}
{Saur}, J., {Grambusch}, T., {Duling}, S., {Neubauer}, F.~M., \& {Simon}, S.
  2013, aap, 552, A119

\bibitem[{{Slavin}(1979)}]{Slavin}
{Slavin}, J.~A. 1979, J. Geophys. Res., 84, 2076

\bibitem[{{Slavin}(2008)}]{Slavin2}
{Slavin}, J.~A. 2008, Science, 321, 85

\bibitem[{{Slavin} {et~al.}(2009){Slavin}, {Acu{\~n}a}, {Anderson}, {Baker},
  {Benna}, {Boardsen}, {Gloeckler}, {Gold}, {Ho}, {Korth}, {Krimigis},
  {McNutt}, {Raines}, {Sarantos}, {Schriver}, {Solomon}, {Tr{\'a}vn{\'{\i}}{\v
  c}ek}, \& {Zurbuchen}}]{2009Sci...324..606S}
{Slavin}, J.~A., {Acu{\~n}a}, M.~H., {Anderson}, B.~J., {et~al.} 2009, Science,
  324, 606

\bibitem[{{Strugarek} {et~al.}(2015){Strugarek}, {Brun}, {Matt}, \&
  {R{\'e}ville}}]{Strugarek}
{Strugarek}, A., {Brun}, A.~S., {Matt}, S.~P., \& {R{\'e}ville}, V. 2015, apj,
  815, 111

\bibitem[{{Varela}(2015)}]{Varela}
{Varela}, J. 2015, Planet. Space Sci., 119, 264

\bibitem[{{Varela}(2016{\natexlab{a}})}]{Varela2}
{Varela}, J. 2016{\natexlab{a}}, Planet. Space Sci., 120, 78

\bibitem[{{Varela}(2016{\natexlab{b}})}]{Varela3}
{Varela}, J. 2016{\natexlab{b}}, Planet. Space Sci., 122, 46

\bibitem[{{Wang} {et~al.}(2010){Wang}, {Mueller}, {Motschmann}, \&
  {Ip}}]{2010Icar..209...46W}
{Wang}, Y.-C., {Mueller}, J., {Motschmann}, U., \& {Ip}, W.-H. 2010, Icarus,
  209, 46

\bibitem[{{Wu}(1979)}]{Wu}
{Wu}, C.~S. 1979, 230, 621

\bibitem[{{Zarka}(1997)}]{Zarka}
{Zarka}, P. 1997, in "Planetary Radio Astronomy IV" (Austrian Acad. Sci. Press:
  Vienna), 101

\bibitem[{{Zarka}(1998)}]{Zarka5}
{Zarka}, P. 1998, J. Geophys. Res., 103, 20159

\bibitem[{{Zarka}(2000)}]{Zarka2}
{Zarka}, P. 2000, "Radio Astronomy at Long Wavelength" Geophysical Monograph,
  119, 167

\bibitem[{{Zarka}(2007)}]{Zarka4}
{Zarka}, P. 2007, Planet. Space Sci., 55, 598

\bibitem[{{Zarka} {et~al.}(2015){Zarka}, {Lazio}, {Hallinan}, \& {SKA/Cradle of
  Life Astrobiology Science Working Group}}]{Zarka6}
{Zarka}, P., {Lazio}, T. J.~W., {Hallinan}, G., \& {SKA/Cradle of Life
  Astrobiology Science Working Group}. 2015, SKA Science Book

\bibitem[{{Zarka} {et~al.}(2001){Zarka}, {Treumann}, {Ryabov}, \&
  {Ryabov}}]{Zarka3}
{Zarka}, P., {Treumann}, R.~A., {Ryabov}, B.~P., \& {Ryabov}, V.~B. 2001,
  Astrophys. Space Sci., 277, 293

\end{thebibliography}

\end{document}